\documentclass[aps,prd,preprint,showpacs,showkeys]{revtex4}

\usepackage{latexsym}

\begin{document}


\title{Entropy of the FRW universe based on the generalized uncertainty
  principle}

\author{Wontae Kim}
\email[]{wtkim@sogang.ac.kr}
\affiliation{Department of Physics,
  Sogang University, Seoul, 121-742, Korea}

\author{Young-Jai Park}
\email[]{yjpark@sogang.ac.kr}
\affiliation{Department of Physics and WCU-SSME Program Division,
Sogang University, Seoul 121-742, Korea}

\author{Myungseok Yoon}
\email[]{younms@sejong.ac.kr}
\affiliation{Institute of Fundamental Physics, Sejong University, Seoul 143-747, Korea}

\date{\today}

\begin{abstract}
  The statistical entropy of the FRW universe described by
  time-dependent metric is newly calculated using the brick wall
  method based on the general uncertainty principle with the minimal
  length. We can determine the minimal length with the Plank scale to
  obtain the entropy proportional to the area of the cosmological
  apparent horizon.
\end{abstract}



\maketitle

\section{Introduction}
\label{sec:intro}

There has been much interest in the holographic principle for
gravity~\cite{ghss,hss, drs,cl,mty}.  It relates the quantum theory of gravity to
quantum field theory without gravity on the boundary with lower
dimensions. As for black holes, the entropy of a black hole is
proportional to the area of the horizon~\cite{bekenstein,bekenstein:1974, hawking}. On
the other hand, it has been claimed that the holographic property on
the entropy also appears in the cosmological context with a particle
horizon~\cite{fs}. Concerning the holography of the FRW universe, the
covariant entropy bound has been described in Ref.~\cite{bousso} and
it has been shown that the $k=1$ FRW universe filled with CFT
(radiation-matter) indicates the holographic nature in terms of
temperature and entropy~\cite{verlinde}. The boundary in the
cosmology can be chosen as the cosmological apparent horizon instead
of the event horizon of a black hole. It can be shown that the
thermodynamic first law can be satisfied and the entropy is $S = A/4G$
with the temperature $T=|\kappa|/2\pi$, where $G$, $A$, and $\kappa$
are the gravitational constant, the area of the apparent horizon, and
the surface gravity on the horizon, respectively~\cite{ck,cc,ac,swc,cc:brane}.

The brick wall method suggested by 't Hooft~\cite{thooft} can be used
for calculating the statistical entropy of a black hole, and the
cutoff parameter is introduced to avoid the divergence near the event
horizon. The method can help us to understand the origin of entropy in
various black holes~\cite{su,jacobson, gm, ao,ao:thermodynamics, dlm, kkps,hkps, mukohyama}.  Since
degrees of freedom of a field are dominant near horizon, the brick
wall model has often been replaced by a thin-layer model, which makes
the calculation of entropy simple~\cite{hzk,lz:film,zl}. Recently, it has been
shown that the (thin-layered) brick wall model can be applied to a
time-dependent black hole with an assumption of local equilibrium near
horizon~\cite{lz}.  It can be also applied to calculation of the
entropy for the Friedmann-Robertson-Walker (FRW)
universe~\cite{ksy,nou}, which is described with time-dependent
metric.  On the other hand, the generalized uncertainty principle
(GUP) modifying the usual Heisenberg's uncertainty principle has been
used in calculating the entropy of various black holes~\cite{li, liu,liu:membrane,
  lhz, sl, kkp,mkp, kp, ko,zrl,li:prd}, where the cutoff parameter is
naturally connected with the minimal length.

In this paper, we would like to study the entropy of the FRW universe
based on the brick wall method considering the GUP, and find the
minimal length giving the entropy proportional to the area. In the
section~\ref{sec:FRW}, we review the FRW cosmology briefly. The
state of the universe will be assumed to be in the locally
thermodynamic equilibrium.  Then, the degrees of freedom of a field
are dominant near horizon like as the case of the black hole so that
the thin layer will be considered as an equilibrium system near the
horizon.  We will calculate the entropy of the FRW universe and
determine the minimal length in terms of the Plank scale and the total
density parameter in the section~\ref{sec:entropy}. As a result, we
will show that the entropy becomes $S = A/4G$ when the cosmological
constant is dominant in compared with the other matters. Some
conclusion and comment are given in section~\ref{sec:dis}.

\section{Background geometry of the FRW universe}
\label{sec:FRW}

The standard metric of the FRW universe is given by
\begin{equation}
  \label{metric}
  ds^2 = - dt^2 + a^2(t) \left[ \frac{dr^2}{1-kr^2} + r^2 d\Omega^2 \right],
\end{equation}
where $a(t)$, $k=0, \pm 1$, and $d\Omega^2 = d\theta^2 +
\sin^2\theta\, d\phi^2$ are the scale factor, the normalized spatial
curvature, and the line element on the unit two-sphere,
respectively. Before we take the brick-wall method to calculate the
entropy of the FRW universe, we need to transform the radial
coordinate to the similar form to the metric of black holes as
follows:
\begin{equation}
  \label{metric:tR}
  ds^2 = \frac{1}{v^2} \left( - f\, dt^2 - 2H R\, dt dR +
    dR^2 \right) + R^2 d\Omega^2,
\end{equation}
where we have defined the new radial coordinate as $R = ar$ and the
Hubble parameter as $H(t) = \dot{a}/a$. The functions $v$ and $f$ are
defined by $v(t,R) = \sqrt{1-kR^2/a^2}$ and $f(t,R) = 1-R^2/R_A^2$
with the apparent horizon of $R_A(t) = 1/\sqrt{H^2 + k/a^2}$. The dot
denotes the derivative with respect to the time coordinate $t$.  
The temperature on the apparent horizon is defined by $T = \beta^{-1}
= \frac{|\kappa|}{2\pi}$ and the surface gravity on the horizon is
given by $\kappa = \frac{1}{2\sqrt{-h}} \partial_a (\sqrt{-h}
h^{ab} \partial_b R)|_{R=R_A}$~\cite{hayward,hma}, where the metric
$h_{ab}$ is defined by $ds^2 = h_{ab} (x) dx^a dx^b + R^2(x)
d\Omega^2$. Then we can write the explicit form of the temperature as
\begin{equation}
  \label{T}
  T = \frac{H^2 R_A}{2\pi} \left| 1 +
    \frac{1}{2H^2} \left( \dot{H} + \frac{k}{a^2} \right) \right|.
\end{equation}

Now, we will review the FRW cosmology briefly. From the
metric~(\ref{metric}), the Einstein equation is given by
\begin{eqnarray}
  H^2 + \frac{k}{a^2} &=& \frac{8\pi G}{3} \rho + \frac13
  \Lambda, \label{FRW:eom1} \\
  \dot{H} - \frac{k}{a^2} &=& -4\pi G(\rho + p), \label{FRW:eom2}
\end{eqnarray}
where $\rho$, $p$, and $\Lambda$ are the energy density, the pressure,
and the cosmological constant, respectively.  The equation-of-state
parameter $\gamma = p/\rho$ is defined by
\begin{equation}
  \label{gamma:def}
  \gamma = \frac{p_{\rm tot}}{\rho_{\rm tot}} = \frac{\Omega_{\rm
      rad}/3 - \Omega_\Lambda}{\Omega_{\rm tot}},
\end{equation}
where $\rho_{\rm tot} = \rho + \rho_\Lambda = \rho_m + \rho_{\rm rad}
+ \rho_\Lambda$ and $p_{\rm tot} = p + p_\Lambda = p_m + p_{\rm rad} +
p_\Lambda$ with $\rho_\Lambda = \Lambda / 8\pi G$ and $\rho_k =
-3k/8\pi G a^2$~\cite{carroll}. In particular, $\gamma$ becomes $1/3$
for the radiation-dominated, zero for the matter($m$)-dominated,
$-1/3$ for the spatial curvature($k$)-dominated, and $-1$ for the
vacuum energy($\Lambda$)-dominated universe.  In general, the
equation-of-state parameter is given by $\gamma = n/3-1$ when an
energy density satisfies the power law, $\rho \sim a^{-n}$.  The
subscript ``$m$'' and ``rad'' mean the matter and the
radiation-dominant, respectively.  The density parameter for the type
of energy $\rho_i$ ($i=m, {\rm rad}, \Lambda, {\rm etc.}$) has been
defined by $\Omega_i = \rho_i / \rho_c$, where $\rho_c = 3H^2/8\pi G =
\rho_{\rm tot} + \rho_k$ is the critical density.  The equations of
motion~(\ref{FRW:eom1}) and (\ref{FRW:eom2}) can be written as
\begin{eqnarray}
  H^2 &=& \frac{8\pi G}{3} (\rho_{\rm tot} +
  \rho_k), \label{FRW:eom1:rho} \\
  \dot{H} + H^2 &=& \frac{4\pi G}{3} (1+3\gamma) \rho_{\rm tot}.
  \label{FRW:eom2:rho}
\end{eqnarray}
Using Eqs.~(\ref{FRW:eom1:rho}) and (\ref{FRW:eom2:rho}), the
temperature~(\ref{T}) is calculated as
\begin{equation}
  \label{T:gamma}
  T = \beta^{-1} = \frac{1}{2\pi R_A} \frac{\left|1-3\gamma\right|}{4}.
\end{equation}
The apparent horizon is also given by
\begin{equation}
  \label{horizon:Omega}
  R_A = \frac{1}{H\sqrt{\Omega_{\rm tot}}},
\end{equation}
where $\Omega_{\rm tot} = 1 - \Omega_k = 1 + k/(H^2 a^2)$. 

\section{Quantum statistical entropy of the FRW universe}
\label{sec:entropy}

Now, let us consider a quantum gas of scalar particles confined within a thin
layer near the apparent horizon of the FRW universe and introduce an
infinitesimal cut-off parameter $\epsilon$.  It is assumed that the
scalar field satisfies the Klein-Gordon equation
\begin{equation}
  \label{KG}
  (\Box - \mu^2) \Phi(t,R,\theta,\phi) = 0,
\end{equation}
with the boundary condition $\Phi(R_A - \epsilon) = \Phi(R_A)$, where
$R_A - \epsilon$ and $R_A$ represent the inner and outer walls of the
layer, respectively, and $\mu$ is the mass of the field. From the
metric~(\ref{metric:tR}), Eq.~(\ref{KG}) can be rewritten as
\begin{equation}
  \label{KG:metric}
  - v \partial_t \left[ \frac{1}{v} (\partial_t \Phi + HR \partial_R
    \Phi) \right] + \frac{v}{R^2} \partial_R \left[ \frac{R^2}{v}
    (-HR \partial_t \Phi + f \partial_R \Phi) \right] + \left(
    \frac{1}{R^2} \nabla_\Omega^2 - \mu^2 \right) \Phi = 0,
\end{equation}
where $\nabla_\Omega^2 \equiv
\csc\theta\ \partial_\theta(\sin\theta\ \partial_\theta) + \csc^2 \theta\
\partial_\phi^2$ means the Laplacian on the unit two sphere. Then, by using
the WKB approximation with $\Phi \sim \exp[i
\sigma(t,R,\theta,\phi)]$, we can obtain the relation as
\begin{equation}
  \label{mom:relation}
  f \, p_R^2 + 2HR\omega \, p_R + \frac{p_\theta^2}{R^2} +
  \frac{p_\phi^2}{R^2\sin^2\theta} = \omega^2 - \mu^2,
\end{equation}
where the energy and the momenta of the scalar field are defined by
$\omega = -\frac{\partial\sigma}{\partial t}$, $p_R =
\frac{\partial\sigma}{\partial R}$, $p_\theta =
\frac{\partial\sigma}{\partial \theta}$, and $p_\phi =
\frac{\partial\sigma}{\partial \phi}$.  Note that
Eq.~(\ref{mom:relation}) can be also obtained from the relation $p_\mu
p^\mu = -\mu^2$ with $p_\mu = (-\omega, p_R, p_\theta, p_\phi)$.
Moreover, we obtain $\vec{p}^2 = p^i p_i = g^{\mu i} p_\mu p_i = f \,
p_R^2 + HR\omega \, p_R + \frac{p_\theta^2}{R^2} +
\frac{p_\phi^2}{R^2\sin^2\theta} = \omega^2 - \mu^2 - HR\omega \, p_R$
and $\frac{p_\theta^2}{R^2} + \frac{p_\phi^2}{R^2\sin^2\theta} =
\omega^2 - \mu^2 - 2HR\omega \, p_R - f \, p_R^2$, from which we have
to read the condition for $p_R$ as $p_R^- \le p_R \le p_R^+$, where
$p_R^\pm$ are defined by $p_R^\pm = \Big[- HR \pm \sqrt{(f+H^2 R^2)
  \omega^2 - \mu^2f} \Big]/f$.

From now on, we will assume such a locally equilibrium system that the
temperature of thermal radiation is slowly varying near the
horizon. The temperature is approximately proportional to the apparent
horizon, $T \sim R_A^{-1}$, which will be shown later, and then, in
the locally equilibrium system it requires that $\delta T/T \sim
\delta R_A / R_A \sim \delta a/a \ll 1$ for $H \ll 1$, where $\delta$
denotes fluctuation of each quantity.

On the other hand, many efforts have been recently devoted to the
generalized uncertainty relation~\cite{kmm,rama,cmot,garay,scardigli,acs}, which is given by
\begin{equation} 
  {\Delta x} {\Delta p} \ge \frac{\hbar}{2}\left[ 1 +
    {\lambda} \left(\frac{{\Delta p}}{\hbar}\right)^2 \right].
\end{equation}
Here $\lambda$ is the GUP parameter, which is related to the proper
length, $2\sqrt{\lambda}$.  It will effectively play a role of the
brick wall cutoff. We simply take the units $\hbar=k_{B}=c \equiv
1$. Then, one can easily get ${\Delta x} \geq \sqrt{\lambda}$, which
gives the lowest bound of the the minimal length near horizon.
Furthermore, based on the generalized uncertainty relation, the
3-dimensional volume of a phase cell is changed from $(2{\pi})^{3}$
into $(2{\pi})^{3}(1 + {\lambda}{\vec{p}^{2}})^{3}$~\cite{kmm,cmot}.

Then, the number of quantum states with the energy less than $\omega$
under the GUP is calculated as
\begin{eqnarray}
  \label{n}
  n(\omega) &=& \frac{1}{(2\pi)^3} \int\, dR d\theta d\phi \frac{dp_R
    dp_\theta dp_\theta}{(1+\lambda \vec{p}^2)^3} \nonumber \\
  &=& \frac{1}{2\pi} \int\, dR R^2 \int_{p_R^-}^{p_R^+} \, dp_R \,
  \frac{\omega^2 - \mu^2 - 2HR\omega\, p_R - f\, p_R^2}{\left[ 1+
      \lambda(\omega^2 - \mu^2 - HR\omega\, p_R) \right]^3} \nonumber
  \\
  &=& \frac{1}{\pi\lambda^2} \int\, dR \Bigg\{ \frac{1}{H^2 \omega^2}
  \frac{ \left[ f + \lambda \left( (f+H^2R^2)\omega^2 - \mu^2 f\right)
    \right] \sqrt{(f+H^2R^2)\omega^2 - \mu^2 f}}{f \left[ 1+
      \lambda(\omega^2 - \mu^2) \right]^2 + \lambda H^2 R^2 \omega^2 \left[ 2+
      \lambda(\omega^2 - \mu^2) \right] } \\
  & & + \frac{f}{2 \lambda H^3 \omega^3 R} \ln \frac{ f+ \lambda
    \left[ (f+H^2 R^2) \omega^2 -\mu^2f - HR\omega \sqrt{(f+H^2
          R^2)\omega^2 - \mu^2 f} \right]}{ f+ \lambda
    \left[ (f+H^2 R^2) \omega^2 -\mu^2f + HR\omega \sqrt{(f+H^2
          R^2)\omega^2 - \mu^2 f} \right]} \Bigg\}. \nonumber
\end{eqnarray}
Since Eq.~(\ref{mom:relation}) can be written
  as $f \left( p_R + \frac{HR}{f} \omega \right)^2 +
  \frac{p_\theta^2}{R^2} + \frac{p_\phi^2}{R^2\sin^2\theta} =
  \left(1+\frac{H^2R^2}{f} \right) \omega^2 - \mu^2$, we require the
  energy condition as $\omega \ge \omega_0$ with $ \omega_0 \equiv
  \mu\sqrt{f}/\sqrt{f+H^2R^2}$. Note that $\omega_0$ goes to zero near
  horizon.

Now, the free energy is given by $F = - \int\, d\omega
\frac{n(\omega)}{e^{\beta\omega}-1}$ with the inverse of temperature
$\beta$, the corresponding entropy is given by
\begin{equation}
  \label{S}
  S = \beta^2 \frac{\partial F}{\partial \beta} = \beta^2 \int\,
  d\omega \frac{\omega\, n(\omega)}{4 \sinh^2 \frac12\beta\omega}.
\end{equation}
For convenience, we replaced $\omega$ by $x \equiv \frac12
\beta\omega$. Note that $f$ and $x_0 = \frac12 \beta\omega_0 = \frac12
\beta\mu\sqrt{f} / \sqrt{f+H^2 R^2}$ vanish while $x_0^2
(1+H^2R^2/f) = \frac14 \beta^2\mu^2$ does not for massive
scalar fields in the near horizon limit of \textit{i.e.}, $R \to R_A$.
Since $f/\lambda$ can be taken as a
finite value near the horizon, the logarithmic term in  Eq.~(\ref{n}) can
be neglected in the leading order approximation.
Then, the entropy~(\ref{S}) becomes
\begin{equation}
  \label{S:I}
  S \approx \frac{\beta^3}{8\pi\lambda^3 H} \int_0^\infty \, dx
  \frac{I(x)}{\sinh^2 x},
\end{equation}
with
\begin{equation}
  \label{I}
  I(x) \equiv \int_{R_A - \epsilon}^{R_A} \, dR\,  \frac{R x^2 + \beta^2 f/(4\lambda H^2
    R)}{x^4 - \frac14 \beta^2 \mu^2 x^2 +
    \beta^2/(2\lambda)}.
\end{equation}
In the leading order of $\epsilon$ , Eq.~(\ref{I}) is calculated as
\begin{equation}
  \label{I:int}
  I(x) = \frac{R_A x^2 \epsilon}{x^4 - \frac14 \beta^2 \mu^2 x^2 +
      \beta^2/(2\lambda) } + O(\epsilon^2).
\end{equation}
Substituting Eq.~(\ref{I:int}) into Eq.~(\ref{S:I}), we obtain
\begin{equation}
  \label{S:x:epsilon}
  S \approx \frac{\beta^3 R_A \epsilon}{8\pi\lambda^3 H} \int_0^\infty
  dx \, h(x),
\end{equation}
where
\begin{equation}
  \label{h}
  h(x) = \frac{x^2}{\sinh^2 x \, \left(x^4 - \frac14 \beta^2 \mu^2 x^2 +
      \frac{\beta^2}{2\lambda} \right) }.
\end{equation}
The entropy (\ref{S:x:epsilon}) can be regarded as the complex
function $h(z) = z^2/[\sinh^2 z \, (z^2 - \alpha_+^2) (z^2 -
\alpha_-^2) ]$ with $\alpha_\pm^2 = \frac18 \beta^2 \mu^2 \pm i
\Gamma^2$ and $\Gamma = \sqrt{\beta/\sqrt{2\lambda}} \left( 1 -
  \beta^2\mu^4 \lambda / 32 \right)^{1/4}$.  If we assume that $\mu^2
\beta < 4\sqrt{2/\lambda}$, then $\Gamma$ is real and positive. The
function $h(z)$ has poles at $z=\pm \alpha_+$, $z=\pm\alpha_-$, and
$z=in\pi$, where $n$'s are non-vanishing integers. The residues of
$h(z)$ are $2\pi n (n^4 \pi ^4 - \alpha_+^2 \alpha_-^2)/[i(n^2 \pi^2 +
\alpha_+)^2 (n^2 \pi^2 + \alpha_-)^2]$ at $z = in\pi$ for $n \ne 0$
and their sum for $n \ge 1$ is obtained as $\pi \lambda /
(i\beta^2)$. The residues of $h(z)$ are $\alpha_\pm \, \mathrm{csch}^2
\alpha_\pm / [2(\alpha_\pm^2 - \alpha_\mp^2)]$ at $z=\alpha_\pm$. At
low temperature, the residues of $h(z)$ at $z^2= \alpha_\pm^2$
decrease exponentially when $\beta$ goes to infinity so that the
leading term of the integral in Eq.~(\ref{S:x:epsilon}) comes from the
poles at $z=in\pi$. Using the residue theorem, the entropy can be
obtained as
\begin{equation}
  \label{S:epsilon}
  S \approx  \frac14 A \cdot \frac{\beta\epsilon}{24\lambda^2 H R_A},
\end{equation}
where $A=4\pi R_A^2$ is the area of the apparent horizon.

Defining a diagonalized metric, $\tilde{g}_{\mu\nu}$ from
Eq.~(\ref{metric:tR}), the proper length is calculated as
\begin{equation}
  \label{lambda:def}
  2\sqrt{\lambda} \equiv \int_{R_A - \epsilon}^{R_A} dR
  \sqrt{\tilde{g}_{RR}} =  \int_{R_A - \epsilon}^{R_A}
  \frac{dR}{\sqrt{f}} \approx \sqrt{2 R_A \epsilon},
\end{equation}
which leads to $\epsilon = 2\lambda/R_A$.
Then, we can obtain the
entropy as follows:
\begin{equation}
  S \approx \frac14 A \cdot \frac{\beta}{12 \lambda H R_A^2} =
  \frac{A}{4} \cdot \frac{\pi}{6\lambda H^3} \left( H^2 +
    \frac{k}{a^2} \right)^{3/2} \left| 1+\frac{1}{2H^2}
    \left( \dot{H} + \frac{k}{a^2} \right) \right|^{-1}, \label{S:H}
\end{equation}
where the temperature~(\ref{T}) has been used.

Now, one can rewrite the above entropy formula in terms of matter
contents using the equations of motion for the FRW cosmology. Inserting
Eqs.~(\ref{T:gamma}) and (\ref{horizon:Omega}) into Eq.~(\ref{S:H})
and recovering the dimension, the entropy becomes
\begin{equation}
  \label{S:energy}
  S \approx \frac{A}{4G} \cdot \frac{2\pi \ell_p^2 \sqrt{\Omega_{\rm
        tot}}}{3\lambda |1-3\gamma|},
\end{equation}
where $\ell_p = \sqrt{G\hbar/c^3}$ is the Plank length.
Setting the minimal length as $2\sqrt{\lambda} = \ell_p \Omega_{\rm
  tot}^{1/4} \sqrt{2\pi/3} \approx 1.4472 \times \Omega_{\rm
  tot}^{1/4} \ell_p$, the entropy is written as
\begin{equation}
  \label{S:delta}
  S \approx \frac{4}{|1-3\gamma|} \frac{A}{4G}.
\end{equation}
According to the recent observational data~\cite{data:observe}, the
density parameter is given by $\Omega_{\rm tot} \simeq 1$ for our
present universe. Therefore, the minimal length is approximately
obtained as $2\sqrt{\lambda} \approx 1.4472 \times \ell_p$.  Note that
the minimal length is exact constant for $k=0$, while it is
time-dependent for $k = \pm 1$.  It is interesting to note that the
vacuum energy($\Lambda$)-dominated universe of $\gamma=-1$ gives the
one quarter area law. 

\section{Conclusions}
\label{sec:dis}

In conclusion, we have newly studied the cosmological entropy applying
the brick wall method based on the GUP to the time-dependent FRW
universe. In fact, there is another way to define the cutoff in that
we could absorb the state parameter, $\gamma$, into the cutoff by
$2\sqrt{\lambda} = \ell_p \Omega_{\rm tot}^{1/4} \sqrt{2\pi/3} /
\sqrt{|1-3\gamma|/4}$ ($\gamma \ne 1/3$), which yields the one quarter
area law. In that case, we can not avoid the era-dependent cutoff.  We
do not deal with the holographic property of the FRW universe in this
paper, but we hope that we will study it in the next work.

The final comment is that for the radiation dominant era of
$\gamma=1/3$, the entropy is divergent. In the present formulation,
the temperature is zero as seen in Eq.~(\ref{T:gamma}), which is
reminiscent of the extremal black hole so that it is not easy to
formulate the free energy and its entropy. We hope this issue will be
discussed in elsewhere. 
 
\begin{acknowledgments}
  W. Kim and Y.-J. Park were supported by the Korea Science and
  Engineering Foundation(KOSEF) grant funded by the Korea
  government(MEST) with grant number 20090083765.  M. Yoon was
  supported by the National Research Foundation of Korea Grant funded
  by the Korean Government [NRF-2009-351-C00109].
\end{acknowledgments}



\end{document}